\documentstyle[12pt]{article}
\newcommand{\es}{\\[2mm]}

\newcommand{\journal}[4]{{\em #1~}#2\,(19#3)\,#4;}

\newcommand{\cmp}{\journal {Comm. Math. Phys.}}

\newcommand{\pl}{\journal {Phys. Lett.}}
\newcommand{\mpl}{\journal {Mod. Phys. Lett.}}
\newcommand{\prep}{\journal {Phys. Reports}}

\newcommand{\annp}{\journal {Ann. Phys. (N.Y.)}}

\makeatletter
\@addtoreset{equation}{section}
\makeatother

\def\Lp{\displaystyle{\biggl(}}
\def\Rp{\displaystyle{\biggr)}}
\def\LP{\displaystyle{\Biggl(}}
 \def\wti{\widetilde}
\def\RP{\displaystyle{\Biggr)}}

\newcommand{\lc}{\left[}\newcommand{\rc}{\right]}

\renewcommand{\a}{\alpha}
\renewcommand{\b}{\beta}
\renewcommand{\d}{\delta}
\newcommand{\e}{\varepsilon}

\renewcommand{\l}{\lambda} 
\newcommand{\lR}{\lambda_{R}}
\newcommand{\lI}{\lambda_{I}}
\newcommand{\m}{\mu}
\newcommand{\n}{\nu}

\renewcommand{\o}{\omega}

\newcommand{\r}{\rho}
\newcommand{\s}{\sigma}

\newcommand{\WT}{\wti T}

\newcommand{\MM}{{\cal M}}

\newcommand{\complex}{{\kern .1em {\raise .47ex
\hbox {$\scriptscriptstyle |$}}
    \kern -.4em {\rm C}}}
\newcommand{\real}{{{\rm I} \kern -.19em {\rm R}}}
\newcommand{\rational}{{\kern .1em {\raise .47ex
\hbox{$\scripscriptstyle |$}}
    \kern -.35em {\rm Q}}}
\renewcommand{\natural}{{\vrule height 1.6ex width
.05em depth 0ex \kern -.35em {\rm N}}}
\newcommand{\dint}{\displaystyle{\int}}
\newcommand{\xint}{\dint d^4 \! x \, }

\newcommand{\pa}{\partial}

\newcommand{\dfrac}[2]{{\displaystyle{\frac{#1}{#2}}}}

\newcommand{\sla}{\raise.15ex\hbox{$/$}\kern -.57em}

\newcommand{\twiddle}{\lower.9ex\rlap{$\kern -.1em\scriptstyle\sim$}}

\newcommand{\vf}{{\varphi}}
\newcommand{\Wvf}{{\wti \varphi}}
\newcommand{\dvf}{{\varphi^{\dagger}}}
\newcommand{\dWvf}{{\wti \varphi}^{\dagger}}

\newcommand{\equ}[1]{(\ref{#1})}
\newcommand{\eq}{\begin{equation}}
\newcommand{\eqn}[1]{\label{#1}\end{equation}}
\newcommand{\eea}{\end{eqnarray}}
\newcommand{\eqa}{\begin{eqnarray}}
\newcommand{\eqan}[1]{\label{#1}\end{eqnarray}}
\newcommand{\ba}{\begin{array}}
\newcommand{\ea}{\end{array}}
\newcommand{\eqac}{\begin{equation}\begin{array}{rcl}}
\newcommand{\eqacn}[1]{\end{array}\label{#1}\end{equation}}

\begin{document}
%********************************************************
{\large     %POUR PREPRINT
%********************************************************
%*********************************************************************
%\def\ftoday{{\sl  \number\day \space\ifcase\month
%\or Janvier\or F\'evrier\or Mars\or avril\or Mai
%\or Juin\or Juillet\or Ao\^ut\or Septembre\or Octobre
%\or Novembre \or D\'ecembre\fi
%\space  \number\year}}
%grav.tex \hspace{1cm} {\bf DRAFT} \hfill{\ftoday}\\

%\hfill{\number\time}\\

%********************************************************
%********************************************************
%PAGE DE TITRE
%*********************************************************
{\ }

\vspace{20mm}

\vspace{2cm}
%***************************************************************
\centerline{\LARGE $\vf_{4}^{4}-$theory for antisymmetric tensor matter}
\vspace{2mm}

\centerline{\LARGE fields in Minkowski space-time}
\vspace{2mm}

%***************************************************************
\vspace{9mm}

\centerline{ Vitor Lemes, Ricardo Renan}
\centerline{ and }
\centerline{S.P. Sorella}
\vspace{4mm}
\centerline{\it C.B.P.F}
\centerline{\it Centro Brasileiro de Pesquisas Fisicas,}
\centerline{\it Rua Xavier Sigaud 150, 22290-180 Urca}
\centerline{\it Rio de Janeiro, Brazil}
\vspace{10mm}

\centerline{{\normalsize {\bf PACS: 11.15.Bt }} }
\vspace{4mm}
\centerline{{\normalsize {\bf REF. 007/95}} }

\vspace{4mm}
\vspace{10mm}

\centerline{\Large{\bf Abstract}}

\vspace{2mm}

\noindent
The nonabelian generalization of a recently proposed abelian axial gauge
model for tensor matter fields is obtained. In both cases
the model can be derived from a $\vf^{4}-$type theory for antisymmetric
fields obeying a complex self-dual condition.
\vspace{2mm}

\setcounter{page}{0}
\thispagestyle{empty}

\vfill
\pagebreak
%\null

%********************************************************
%********************************************************
\section{Introduction}

In a recent paper L. V. Avdeev and M. V. Chizhov~\cite{ac}
have analysed
the properties of an abelian axial gauge model containing antisymmetric second
rank tensor fields as matter fields. The model is formulated in Minkowski
flat space-time and exihibits several interesting features which allow for
many applications both from phenomenological~\cite{mvc} and
theoretical~\cite{mvc1} point of view.

Let us underline, in particular, the asymptotically free ultraviolet behaviour
of the abelian axial gauge interaction. As shown by the authors~\cite{ac}
with an
explicit one-loop computation, this is due to the fact that the contribution
of the tensor fields to the gauge $\beta-$function is negative. This
particularly
attractive feature motivates further efforts in understanding tensor matter
fields.

The aim of this work, which is the continuation of a previous paper~\cite{lrs}
where the renormalizability of the model has been discussed to all orders of
perturbation theory, is to study the geometrical properties of
the matter tensor action, i.e. to investigate
the guide principle which underlies its gauge formulation.
In particular, it turns out that the model can be obtained in a very simple and
suggestive way from a $\vf^4-$theory for tensor fields satisfying a complex
self-dual condition, namely
\eq
   \vf_{\m\n} = i \Wvf_{\m\n} \ , \qquad
   \Wvf_{\m\n} = \dfrac{1}{2}\e_{\m\n\r\s}\vf^{\r\s} \ ,
\eqn{complex-self-dual}
$\vf_{\m\n}$ being an antisymmetric complex tensor field and $\e_{\m\n\r\s}$
the
Levi-Civita symbol.

As we shall see in detail, the complex self-dual condition
\equ{complex-self-dual}
uniquely fixes the Lorentz contractions of the tensor $\vf^4-$Lagrangian,
reproducing
thus the action of Avdeev and Chizhov. Moreover, the formulation of the model
as a kind of $\vf^4-$theory will give us a straightforward way of obtaining
its nonabelian generalization which, so far, has not yet been established.

Let us also remark that complex self-dual conditions of the type of
 eq.\equ{complex-self-dual} have been known and used since several years.
As an example we mention the complex self-dual connection\footnote{In this case
the complex self-dual condition refers to the internal indices of $SO(3,1)$
rather than space-time indices.} used by A. Ashtekar in its formulation
of gravity~\cite{ash}.

The paper is organized as follows. In Sect.2, after introducing the complex
self-dual condition \equ{complex-self-dual}, we show how it can be used
to recover the abelian model of Avdeev and Chizhov.
Sect.3 is then devoted to a detailed discussion of the nonabelian
generalization.
Concerning the latter case, we shall limit here only to classical aspects. The
renormalization properties of the nonabelian model, i.e. its stability under
radiative corrections and the absence of anomalies, will be reported in a
future work.

%********************************************************
%********************************************************
\section{Tensor matter fields: the abelian case}

Let us begin this section by introducing the notations and some properties of
the
Levi-Civita tensor $\e_{\m\n\r\s}$. We shall work in a flat Minkowski
space-time
with metric $g_{\m\n}=diag(+,-,-,-)$.

The totally antisymmetric tensor $\e_{\m\n\r\s}$ is normalized as~\cite{itz}
\eq
     \e_{1234}=1 \ , \qquad \e^{1234}= - 1 \ .
\eqn{e-tensor-norm}
It obeys the property
\eq
    \e_{\m_{1}\m_{2}\m_{3}\m_{4}} \e^{\n_{1}\n_{2}\n_{3}\n_{4}} =
      - \d^{[\n_{1}}_{\m_{1}}... \d^{\n_{4}]}_{\m_{4}}  \ ,
\eqn{eps-normal}
from which it follows that
\eq
     \e^{\m\n\r\s} \e_{\r\s\tau\o} =
    -2 ( \d^{\m}_{\tau} \d^{\n}_{\o} - \d^{\m}_{\o} \d^{\n}_{\tau} ) \ .
\eqn{delta-norm}

\subsection{Complex self-dual fields in Minkowski space-time}

A well known feature of the Minkowski space-time is that it does not allow for
self-dual (or antiself-dual) fields
\eq
   \eta_{\m\n} = {\wti \eta}_{\m\n} = \dfrac{1}{2}\e_{\m\n\r\s}\eta^{\r\s} \ .
\eqn{self-dual-fields}
Indeed, due to eq.\equ{delta-norm}, one has
\eq
  {\wti {\wti \eta}}_{\m\n} = - \eta_{\m\n}   \ ,
\eqn{incomp}
which is incompatible with the self-duality condition \equ{self-dual-fields}.

Instead, equation \equ{self-dual-fields} is replaced by a complex self-dual
condition involving a complex tensor field  $\vf_{\m\n}$:
\eq
   \vf_{\m\n} = i \Wvf_{\m\n} \ .
\eqn{c-self-dual}
The factor $i$ in eq.\equ{c-self-dual} is needed in order to compensate the
minus
sign coming from eq.\equ{delta-norm}. In fact, making the dual of
eq.\equ{c-self-dual} one has
\eq
     \Wvf_{\m\n} = i {\wti \Wvf}_{\m\n} = - i \vf_{\m\n}  \ ,
\eqn{consistent-cond}
i.e. one consistently goes back to eq.\equ{c-self-dual}.

The complex self-dual condition  \equ{c-self-dual} is easily solved. Writing
$\vf_{\m\n}$ as
\eq
 \vf_{\m\n} = T_{\m\n} + i R_{\m\n} \ ,
\eqn{T-R-dec}
with $T$ and $R$ real antisymmetric fields, one gets $R= {\wti T}$, i.e.
\eq
 \vf_{\m\n} = T_{\m\n} + i {\wti T}_{\m\n} \ .
\eqn{c-self-dual-sol}

\subsection{Coupling to abelian gauge fields}

Let us now try to couple the complex self-dual field $\vf_{\m\n}$, considered
as a
matter field, to a gauge potential $A_{\m}$. In order to do this we require
that
under an abelian gauge transformation
\eq
\d A_\m = \pa_\m\a  \ ,
\eqn{ab-gauge-transf}
the antisymmetric field $\vf_{\m\n}$ transforms as an ordinary  matter field
according to
\eq
      \d \vf_{\m\n} = i \a  \vf_{\m\n} \ , \qquad
      \d \dvf_{\m\n} = - i \a  \dvf_{\m\n} \ .
\eqn{vf-gauge-transf}
As usual, for the covariant derivative we get
\eq
 \nabla_{\s} \vf_{\m\n} = \pa_{\s} \vf_{\m\n}  - i A_{\s} \vf_{\m\n} \ ,
\eqn{cov-der}
and
\eq
 \d (\nabla_{\s}\vf_{\m\n})  = i \a  (\nabla_{\s}\vf_{\m\n}) \ , \qquad
 \d { (\nabla_{\s}\vf_{\m\n}) }^{\dagger}  = - i \a
    { (\nabla_{\s}\vf_{\m\n}) }^{\dagger} \ .
\eqn{cov-der-gauge-transf}
The next natural step is then to discuss the action. In order to find such an
invariant action let us forget, for the time being, the Lorentz
structure of the
antisymmetric field $\vf$. It is apparent thus that an action
of the $\vf^4-$type theory yields an invariant action:
\eq
 S = \xint \Lp \, {(\nabla \vf)}^{\dagger} (\nabla \vf) - \dfrac{q}{8}
{(\vf^{\dagger}\vf)}^2 \, \Rp \ .
\eqn{f4-type-action}
Let us now try to take into account the Lorentz indices of the field
$\vf_{\m\n}$
and of the covariant derivative $\nabla_{\s}$. Owing to the tensorial nature of
$\vf_{\m\n}$ one could expect many possible Lorentz contractions, both in the
kinetic and in the quartic self-interaction term. Of course, this would spoil
the
interest and the meaning of the action \equ{f4-type-action}.

However, it is a remarkable fact that the complex self-dual condition
\equ{c-self-dual} completely fixes the Lorentz structure of
\equ{f4-type-action},
giving rise to a unique term both in the kinetic and in the self-interaction
sector. In other words, condition \equ{c-self-dual} singles out a unique
invariant action. This nice feature is due to the following property
\eq
      \vf^{\m\n} \MM \vf_{\m\n}^{\dagger} = 0 \ ,
\eqn{chiral-condition}
$\MM$ denoting an arbitrary operator depending on the gauge potential $A_{\m}$
and
on the space-time derivatives $\pa_{\m}$. As one can easily understand,
eq.\equ{chiral-condition} is a direct consequence of the complex self-dual
condition \equ{c-self-dual} and of eq.\equ{delta-norm}. Indeed
\eq\ba{rl}
 \vf^{\m\n} \MM \vf_{\m\n}^{\dagger} & =  \Wvf^{\m\n} \MM \Wvf_{\m\n}^{\dagger}
       = \dfrac{1}{4} \e^{\m\n\a\b} \e_{\m\n\lambda\d}
         \vf_{\a\b} \MM {\vf^{\dagger\lambda\d}} \es
  & = -  \vf^{\lambda\d} \MM \dvf_{\lambda\d}  \ .
\ea\eqn{proof-chiral-cond}
It is worth to notice that condition \equ{chiral-condition} forbids the
existence of a mass term $\vf^{\m\n} \dvf_{\m\n}$, i.e. $(\MM=1)$.

Concerning now the kinetic part of the action \equ{f4-type-action} it is easily
verified that the unique nonvanishing Lorentz contraction is given, modulo
integrations by parts, by
\eq
 \xint \, (\nabla_{\m} \vf^{\m\n}) {(\nabla_{\s} \vf^{\s}_{{\ }\n})}^{\dagger}
\ .
\eqn{kin-Lorentz-contr}
Let us now turn to the quartic self-interaction part. In this case
eq.\equ{chiral-condition} selects three possible terms, given respectively by
\eq
  i) {\ } {\vf^{\dagger\m\n}}{\dvf_{\m\n}} \vf^{\a\b} \vf_{\a\b} \ ,
 {\ }{\ }{\ }
 ii) {\ } {\vf^{\dagger\m\n}} \vf_{\n\a} {\vf^{\dagger\a\b}} \vf_{\b\m} \ ,
 {\ }{\ }{\ }
 iii) {\ }{\vf^{\dagger\m\n}} {\dvf_{\n\a}} \vf^{\a\b} \vf_{\b\m} \ .
\eqn{three-terms}
However, it turns out that these three terms are, in fact, equivalent. This
can be proven by making use of the following identity, valid in four dimension
\eq
 \e_{\a\b\m\n} \Xi_{\s...} +  \e_{\s\a\b\m} \Xi_{\n...} +
 \e_{\n\s\a\b} \Xi_{\m...} +  \e_{\m\n\s\a} \Xi_{\b...} +
 \e_{\b\m\n\s} \Xi_{\a...} = 0 \ ,
\eqn{four-dim-identity}
where $\Xi_{\m...}$ denotes an arbitrary tensor. Equation
\equ{four-dim-identity}
stems from the fact that in four dimension the antisymmetrization with
respect to five Lorentz indices automatically vanishes.

As an example, let us prove the equivalence between the terms $i)$ and $iii)$
of
\equ{three-terms}. We have
\eq
  {\vf^{\dagger\m\n}}{\dvf_{\m\n}} \vf^{\a\b} \vf_{\a\b} =
    -i {\vf^{\dagger\m\n}}{\dWvf_{\m\n}} \vf^{\a\b} \vf_{\a\b} =
    - \dfrac{i}{2}
  \e_{\m\n\lambda\s}{\vf^{\dagger\m\n}}{\vf^{\dagger\lambda\s}}
    \vf^{\a\b} \vf_{\a\b} \ .
\eqn{proof-i-iii}
Moreover, antisymmetrization with respect to the indices $(\m,\n,\lambda,\s)$
and $\a$ yields
\eq\ba{rl}
 {\vf^{\dagger\m\n}}{\dvf_{\m\n}} \vf^{\a\b} \vf_{\a\b}  = & {\ }{\ }
  \dfrac{i}{2}{\vf^{\dagger\m\n}}{\vf^{\dagger\lambda\s}}\vf^{\a\b}
  \Lp \e_{\a\m\n\lambda}\vf_{\s\b} + \e_{\s\a\m\n}\vf_{\lambda\b} \Rp   \es
 & + \dfrac{i}{2}{\vf^{\dagger\m\n}}{\vf^{\dagger\lambda\s}}\vf^{\a\b}
   \Lp   \e_{\lambda\s\a\m}\vf_{\n\b} + \e_{\n\lambda\s\a}\vf_{\m\b} \Rp \es
  = & 4 \,  {\vf^{\dagger\m\n}} {\dvf_{\n\a}} \vf^{\a\b} \vf_{\b\m}  \ ,
\ea\eqn{final-proof-i-iii}
showing then the equivalence.

Summarizing, the complex self-dual condition \equ{c-self-dual} uniquely
fixes the Lorentz structure of the invariant action \equ{f4-type-action}.
The latter, including also the Maxwell term, is given by
\eq\ba{rl}
S_{inv} = & -\dfrac{1}{4g^2} \xint F_{\m\n} F^{\m\n} \\[4mm]
 & - \xint  \Lp \, (\nabla_{\m} \vf^{\m\n})
      {(\nabla_{\s} \vf^{\s}_{{\ }\n})}^{\dagger}
  + \dfrac{q}{8} ( {\vf^{\dagger\m\n}} \vf_{\n\a} {\vf^{\dagger\a\b}}
         \vf_{\b\m}) \, \Rp \ .
\ea\eqn{abelian-inv-action}

\subsection{The Avdeev-Chizhov abelian action}

In order to recover the action of Avdeev and Chizhov~\cite{ac},
let us rewrite the
expression \equ{abelian-inv-action} in components, i.e. let us make use
of eq.\equ{c-self-dual-sol} $(\vf = T + i {\wti T} )$. Concerning the gauge
transformations \equ{ab-gauge-transf}, \equ{vf-gauge-transf}, they split as
\eq
   \d A_{\m}= \pa_{\m}\a \ , \qquad
   \d T_{\m\n} = -\a \WT_{\m\n} \ , \qquad
   \d \WT_{\m\n} = \a T_{\m\n} \ .
\eqn{ab-gauge-T-transf}
Analogously, for the covariant derivative $\nabla_{\s} \vf_{\m\n}$ we get
\eq
 \nabla_{\s} \vf_{\m\n} = \nabla_{\s} T_{\m\n} + i  \nabla_{\s} \WT_{\m\n} \ ,
\eqn{dec-cov-der}
with
\eq
  \nabla_{\s} T_{\m\n} =  \pa_{\s} T_{\m\n} + A_{\s} \WT_{\m\n} \ , \qquad
  \nabla_{\s} \WT_{\m\n} = \pa_{\s} \WT_{\m\n} -  A_{\s} T_{\m\n} \ ,
\eqn{T-WT-cov-der}
and
\eq
  \d (\nabla_{\s} T_{\m\n}) = -\a (\nabla_{\s} \WT_{\m\n}) \ , \qquad
  \d (\nabla_{\s} \WT_{\m\n}) = \a (\nabla_{\s} T_{\m\n}) \ .
\eqn{gauge-tr-T-WT-cov-der}
We recover then the two covariant derivatives $(\nabla T, \nabla\WT)$ already
introduced in~\cite{lrs}.

Finally, after a straightforward calculation, for the invariant action we get
\eq\ba{rl}
S_{inv} = & -\dfrac{1}{4g^2} \xint F_{\m\n} F^{\m\n} \\[4mm]
 & - \xint  \LP \, (\nabla_{\m} T^{\m\n}) (\nabla_{\s} T^{\s}_{{\ }\n})
               +   (\nabla_{\m} \WT^{\m\n}) (\nabla_{\s} \WT^{\s}_{{\ }\n}) \es
 & \qquad \qquad + \dfrac{q}{4} \Lp\,  2 T_{\m\n}T^{\n\r}T_{\r\l}T^{\l\m}
                         - \dfrac{1}{2}{(T_{\m\n}T^{\m\n})}^2 \, \Rp \, \RP
\\[4mm]
 = &  -\dfrac{1}{4g^2} \xint F_{\m\n} F^{\m\n} \\[4mm]
 & + \xint \LP \, \dfrac{1}{2} {(\pa_{\l}T_{\m\n})}^2  - 2
{(\pa_{\m}T^{\m\n})}^2
        + 2 A_{\m}\Lp T^{\m\n}\pa_{\l}\WT^{\l}_{{\ }\n} -
                   \WT^{\m\n}\pa_{\l} T^{\l}_{{\ }\n} \Rp  \es
& \qquad \qquad
        + \Lp \dfrac{1}{2} {(A_{\l}T_{\m\n})}^2 -2 {(A^{\m}T_{\m\n})}^2 \Rp \es
 & \qquad \qquad - \dfrac{q}{4} \Lp \,  2 T_{\m\n}T^{\n\r}T_{\r\l}T^{\l\m}
                         - \dfrac{1}{2}{(T_{\m\n}T^{\m\n})}^2 \, \Rp \, \RP \ ,
\ea\eqn{avd-chi-inv-action}
with
\eq  \d S_{inv}= 0 \ .
\eqn{inv-abelian-cond}
Expression \equ{avd-chi-inv-action} is nothing but the original abelian action
proposed by Avdeev and Chizhov~\cite{ac}.
We see thus that, as announced, the model can
be derived in a very simple and suggestive way from a $\vf^4-$theory for tensor
fields obeying a complex self-dual condition.

\section{The nonabelian case}

In order to obtain the nonabelian generalization of the invariant action
\equ{avd-chi-inv-action}, we proceed as before and we treat the antisymmetric
tensor field $\vf$ as an ordinary bosonic matter field belonging to some
finite representation $(\l^a)^{ij}$ of a Lie group $G$, assumed to be
semisimple (the index $a$ labels the generators of $G$, while the indices
$(ij)$ specify the representation).
For reasons which will be clear later on, the representation identified by the
hermitian matrices $(\l^a)^{ij}$ will be required to be a complex
representation, i.e.
\eq
   \l^a = \lR^a + i  \lI^a  \ ,
\eqn{complex-repres}
$\lR^a$ and $\lI^a$ denoting respectively the real and the imaginary part
of $\l^a$. In particular, from the commutation relations
\eq
   \lc \l^a, \l^b \rc = i f^{abc} \l^c  \ ,
\eqn{generators-comm}
we get
\eq\ba{l}
 \lc \lR^a, \lR^b \rc - \lc \lI^a, \lI^b \rc =  - f^{abc} \lI^c  \ , \es
 \lc \lR^a, \lI^b \rc + \lc \lI^a, \lR^b \rc =  \, f^{abc} \lR^c \ ,
\ea\eqn{lI-lR-relations}
and from the hermiticity condition $\l^a = \l^{a\dagger}$
\eq
  (\lR^a)^{ij} =  (\lR^a)^{ji} \ , \qquad (\lI^a)^{ij} =  - (\lI^a)^{ji} \ .
\eqn{herm-condition}

\subsection{Coupling to Yang-Mills fields}

As said before, in order to couple the nonabelian complex self-dual field
$\vf_{\m\n}^i$
\eq
  \vf_{\m\n}^i = i \Wvf_{\m\n}^i  \ , \qquad
  \vf_{\m\n}^i = T_{\m\n}^i + i \WT_{\m\n}^i  \ ,
\eqn{nonabelian-tensor-field}
to Yang-Mills fields $A^{a}_{\m}$, we treat it as a bosonic matter
field which transforms according to the usual nonabelian gauge transformations,
here written for convenience as $BRS$ transformations~\cite{brs,ol}
\eq\ba{l}
  s A^{a}_{\m} = \pa_{\m}c^a +  f^{abc}  A^{b}_{\m} c^c \ , \es
  s \vf_{\m\n}^i = i c^a  (\l^a)^{ij}\vf_{\m\n}^j \ ,  \es
  s \dvf_{\m\n}^i = - i  c^a \dvf_{\m\n}^j (\l^a)^{ji} \ , \es
  s c^a = - \dfrac{1}{2}  f^{abc} c^b c^c \ , \qquad s^2=0 \ .
\ea\eqn{brs-transf}
In complete analogy with the previous abelian case, for the nonabelian
covariant
derivative $(\nabla_{\s} \vf_{\m\n})^i$ we get
\eq
 (\nabla_{\s} \vf_{\m\n})^i =
       \pa_{\s} \vf_{\m\n}^i  - i A_{\s}^a  (\l^a)^{ij}\vf_{\m\n}^j  \ ,
\eqn{nonabelian-cov-der}
and
\eq\ba{l}
  s (\nabla_{\s} \vf_{\m\n})^i = \, i c^a  (\l^a)^{ij} (\nabla_{\s}
\vf_{\m\n})^j
     \ , \es
  s (\nabla_{\s} \vf_{\m\n})^{\dagger i} = - i c^a
           (\nabla_{\s} \vf_{\m\n})^{\dagger j}  (\l^a)^{ji} \ .
\ea\eqn{brs-transf-nonab-covder}
Of course, property \equ{chiral-condition} remains unchanged, implying thus the
following expression for the $BRS$ invariant nonabelian action:
\eq\ba{rl}
S_{inv} = & -\dfrac{1}{4g^2} \xint F^a_{\m\n} F^{a\m\n} \\[4mm]
 & - \xint  \Lp \, (\nabla_{\m} \vf^{\m\n})^i
      {(\nabla_{\s} \vf^{\s}_{{\ }\n})}^{\dagger i}
  + \dfrac{q}{8}
   ( {\vf^{\dagger\m\n i}} \vf^i_{\n\a} {\vf^{\dagger\a\b j}} \vf^j_{\b\m})
       \, \Rp \ ,
\ea\eqn{nonabelian-inv-action}
where the Yang-Mills term has been included.

\subsection{Components}

For a better understanding of the above nonabelian invariant action let us
rewrite,
as done before, the expression \equ{nonabelian-inv-action} in terms of
the component fields $(T_{\m\n}^i,  \WT_{\m\n}^i)$ of
eq.\equ{nonabelian-tensor-field}. Considering first the $BRS$ transformations
\equ{brs-transf}, we obtain
\eq\ba{l}
  s A^{a}_{\m} = \pa_{\m}c^a +  f^{abc}  A^{b}_{\m} c^c \ ,  \qquad
  s c^a = - \dfrac{1}{2}  f^{abc} c^b c^c \es
  s T_{\m\n}^i = - c^a  \Lp (\lR^a)^{ij}\WT_{\m\n}^j + (\lI^a)^{ij}T_{\m\n}^j
\Rp
               \ ,  \es
  s \WT_{\m\n}^i = \, c^a
     \Lp (\lR^a)^{ij}T_{\m\n}^j - (\lI^a)^{ij}\WT_{\m\n}^j \Rp
   \ .
\ea\eqn{T-WT-brs-transf}
Their nilpotency easily follows from the algebraic relations
\equ{lI-lR-relations}.

One should notice that the choice of a complex representation, i.e. $\lR^a \ne
0$,
allows for a nontrivial mixing between the chiral components $(T,\WT)$ of
the complex self-dual field $\vf$, yielding then the nonabelian generalization
of the Avdeev-Chizhov chiral transformations \equ{ab-gauge-T-transf}.

For the covariant derivative \equ{nonabelian-cov-der} we get
\eq
 (\nabla_{\s} \vf_{\m\n})^i = (\nabla_{\s} T_{\m\n})^i
        + i  (\nabla_{\s} \WT_{\m\n})^i  \ ,
\eqn{nonabelian-dec-cov-der}
with
\eq\ba{l}
(\nabla_{\s} T_{\m\n})^i = \pa_{\s} T_{\m\n}^i
    + A_{\s}^a (\lI^a)^{ij} T_{\m\n}^j  + A_{\s}^a (\lR^a)^{ij} \WT_{\m\n}^j
    \ ,          \es
(\nabla_{\s} \WT_{\m\n})^i = \pa_{\s} \WT_{\m\n}^i
    + A_{\s}^a (\lI^a)^{ij} \WT_{\m\n}^j - A_{\s}^a (\lR^a)^{ij} T_{\m\n}^j \ .
\ea\eqn{T-WT-nonabelian-cov-der}
Their $BRS$ transformations read
\eq\ba{l}
 s (\nabla_{\s} T_{\m\n})^i = - c^a \LP
 (\lR^a)^{ij} (\nabla_{\s} \WT_{\m\n})^j + (\lI^a)^{ij}(\nabla_{\s} T_{\m\n})^j
\Rp
 \ , \es
 s (\nabla_{\s} \WT_{\m\n})^i = \, c^a \LP
 (\lR^a)^{ij} (\nabla_{\s} T_{\m\n})^j - (\lI^a)^{ij}(\nabla_{\s} \WT_{\m\n})^j
\Rp
\ .
\ea\eqn{T-WT-cov-der-BRS-transf}
Finally, for the invariant action \equ{nonabelian-inv-action}
 one has
\eq\ba{rl}
S_{inv} = & -\dfrac{1}{4g^2} \xint F^a_{\m\n} F^{a\m\n} \\[4mm]
 & - \xint  \LP \, (\nabla_{\m} T^{\m\n})^i (\nabla_{\s} T^{\s}_{{\ }\n})^i
               +  (\nabla_{\m} \WT^{\m\n})^i (\nabla_{\s} \WT^{\s}_{{\ }\n})^i
\es
 & \qquad \qquad + \dfrac{q}{4} \Lp\,  2 (T^i_{\m\n}T^{i\n\r})^2
            - \dfrac{1}{2}{(T^i_{\m\n}T^{i\m\n})}^2 \, \Rp \, \RP \\[4mm]
 = &  -\dfrac{1}{4g^2} \xint F^a_{\m\n} F^{a\m\n}
  + \xint \LP \, \dfrac{1}{2} {(\pa_{\l}T_{\m\n})}^2 - 2 {(\pa_{\m}T^{\m\n})}^2
  \RP   \\[4mm]
 & - 2 \xint   A^a_{\m}\Lp (\pa_{\s}T^{\s\n}) \lR^a \WT^{\m}_{{\ }\n} -
                        (\pa_{\s}\WT^{\s\n}) \lR^a T^{\m}_{{\ }\n} \Rp \\[4mm]
&  - 2 \xint   A^a_{\m}\Lp (\pa_{\s}T^{\s\n}) \lI^a T^{\m}_{{\ }\n} +
                        (\pa_{\s} T^{\m\n}) \lI^a T^{\s}_{{\ }\n} \Rp \\[4mm]
&  + \xint A^a_{\m} A^b_{\s} \Lp
     T^{\m\n} \lI^a  \lI^b T^{\s}_{{\ }\n} +
     T^{\m\n} \lI^a  \lR^b \WT^{\s}_{{\ }\n} \Rp \\[4mm]
& - \xint  A^a_{\m} A^b_{\s} \Lp
     \WT^{\m\n} \lR^a  \lI^b T^{\s}_{{\ }\n} +
     \WT^{\m\n} \lR^a  \lR^b \WT^{\s}_{{\ }\n} \Rp  \\[4mm]
&  + \xint A^a_{\m} A^b_{\s} \Lp
     \WT^{\m\n} \lI^a  \lI^b \WT^{\s}_{{\ }\n} -
     \WT^{\m\n} \lI^a  \lR^b T^{\s}_{{\ }\n} \Rp \\[4mm]
& + \xint  A^a_{\m} A^b_{\s} \Lp
     T^{\m\n} \lR^a  \lI^b \WT^{\s}_{{\ }\n} -
     T^{\m\n} \lR^a  \lR^b T^{\s}_{{\ }\n} \Rp  \\[4mm]
& - \dfrac{q}{4} \xint  \Lp\,  2 (T_{\m\n}T^{\n\r})^2
            - \dfrac{1}{2}{(T_{\m\n}T^{\m\n})}^2 \, \Rp \ ,
\ea\eqn{avdeev-chizhov-nonabelian-action}
where the implicit notation
$T \lR^a \WT = T^{i} (\lR^a)^{ij} \WT^{j}$, etc...., has been used.

Expression \equ{avdeev-chizhov-nonabelian-action} represents thus the
nonabelian generalization of the Avdeev-Chizhov model. In particular one
remarks that, contrary to the abelian case \equ{avd-chi-inv-action}, the
Levi-Civita tensor $\e_{\m\n\r\s}$ is now present also in the quartic $(AATT)$
interaction term.

\section{Conclusion}

The abelian axial model proposed by Avdeev and Chizhov has been proven to be
interpreted as a $\vf^4-$type theory for tensor fields obeying a complex
self-dual condition. This formulation allows us to obtain in a simple and
elegant way the corresponding nonabelian generalization.

Many aspects of the tensor matter field theories remain still to
be discussed and
clarified. Exemples of them are, for instance, the unitarity properties of
the degrees of freedom associated to the antisymmetric tensor fields (see also
ref~\cite{mvc1} for a study of the Fock space) and the renormalizability of the
nonabelian model. Any progress on these aspects will be reported in a
detailed work.

\vspace{2cm}

\noindent{\large{\bf Acknowledgements}}

We wish to thank Prof. J. J. Giambiagi and Prof.
Jos{\'e} Helayel-Neto for helpful discussions.

The {\it Conselho Nacional de Desenvolvimento Cientifico e Tecnologico},
$CNPq$-Brazil is gratefully acknowledged for the financial support.

%********************************************************************

}    %POUR PREPRINT
%*********************************************************************
\end{document}